\documentclass[aps,pra,twocolumn,showpacs,superscriptaddress,10pt]{revtex4-1}  
\usepackage{graphicx}  
\usepackage{dcolumn}   
\usepackage{bm}        
\usepackage{amssymb}   
\usepackage{color}				 
\usepackage{amsmath}   
\usepackage{soul}

\begin{document}

\title{Linear-Optic Heralded Photon Source}
\date{\today}

\author{Thiago Ferreira da Silva}\email{thiago@opto.cetuc.puc-rio.br}
\affiliation{Center for Telecommunications Studies, Pontifical Catholic University of Rio de Janeiro, 22451-900, Rio de Janeiro, Brazil}
\affiliation{Optical Metrology Division, National Institute of Metrology, Quality and Technology, 25250-020, Duque de Caxias, Brazil}

\author{Gustavo C. Amaral}
\affiliation{Center for Telecommunications Studies, Pontifical Catholic University of Rio de Janeiro, 22451-900, Rio de Janeiro, Brazil}

\author{Guilherme P. Tempor\~{a}o}
\affiliation{Center for Telecommunications Studies, Pontifical Catholic University of Rio de Janeiro, 22451-900, Rio de Janeiro, Brazil}

\author{Jean Pierre von der Weid}
\affiliation{Center for Telecommunications Studies, Pontifical Catholic University of Rio de Janeiro, 22451-900, Rio de Janeiro, Brazil}

\begin{abstract}
We present a Heralded Photon Source based only on linear optics and weak coherent states. By time-tuning a Hong-Ou-Mandel interferometer fed with frequency-displaced coherent states, the output photons can be synchronously heralded following sub-Poisson statistics, which is indicated by the second-order correlation function ($g^2\left(0\right)=0.556$). The absence of phase-matching restrictions makes the source widely tunable, with 100-nm spectral tunability on the telecom bands. The technique presents yield comparable to state-of-the-art spontaneous parametric down-conversion-based sources, with high coherence and fiber-optic quantum communication compatibility.
\end{abstract}

\maketitle

\section{Introduction}
\label{chap1}

Single-photon sources are a fundamental resource in quantum optics, with a broad range of applications \cite{TanzilliNature05,KnillNature01,GisinNP07,GisinRMP02}. Even though deterministic single photon generation is highly desirable, it is not yet available. Therefore, the photon number statistics of any pseudo-single-photon source is always a concern, and the compromise between vacuum, single- and multi-photon emission probabilities directly impacts the source's performance. 

A practical and low-cost way to probabilistically producing single-photon pulses is to use a faint laser source, whose output pulses approach coherent states \cite{KnightBOOK}. Each pulse is described as a superposition of Fock states with the number of photons following Poisson distribution according to the parameter $\mu$, the average number of photons per pulse. A small value of $\mu$ binds the proportion of multi-photon to non-vacuum pulses to $\mu/2$ but leads to an increase in the probability of emitting vacuum pulses to about $1-\mu$. The weak coherent states (WCSs) lead to a trade-off, as multi-photon and vacuum pulses impose limitations to the security and performance of quantum key distribution (QKD) links. Safeguarded by strict security proofs \cite{GottesmanQIC04,LutkenhausPRA00} and techniques \cite{HwangPRL03,MaPRL05,WangPRL05}, however, WCSs are widely used in QKD.

Alternatively to a faint laser source, nonlinear optical processes can be explored to approach a single-photon source with sub-Poisson photon-number distribution, which is the case of Spontaneous Parametric Down Conversion (SPDC) \cite{HongPRA85,HongPRL86,KwiatPRL95} and Four-Wave Mixing \cite{FiorentinoPTL02,RarityOPEX05}. In such processes, pump photons can be converted into pairs of highly-correlated phase-matched photons. The detection of one photon can, thus, be used to announce the presence of the other photon. This so-called heralded single-photon source can be obtained from continuous-wave (CW) \cite{FaselNJP04,HalderNJP08,WangPRL08} or pulsed \cite{PomaricoOPEX12,NgahLPR15} optical pump even though the inherent low-coherence nature of the non-linear processes yield broadband emission, with $\sim$ 10 ps coherence time \cite{HalderNJP08}. This feature renders the synchronization of fiber-optic QKD systems critical -- specially if it involves the joint measure of two single-photon pulses from distinct remote sources (e.g. Bell-state measurements \cite{PanRMP2012,ThiagoPRA2013,ThiagoJLT13}). The coherence time of the photons can be enhanced to the ns range by using ultra-narrow filtering techniques, for which delicate thermal stabilization is required \cite{HalderNJP08, ClausenNJP14}. Nevertheless, the narrower the filtering, the less bright is the emission and, thus, the throughput of the source. Other rather complex resources, such as double-resonant cavities, could also be employed to achieve higher coherence \cite{ClausenNJP14}. These advanced techniques, however, are frequency-dependent providing barely tunable emission.

In this paper, we propose and experimentally demonstrate a heralded photon source (HPS) fully based on WCSs and completely independent of nonlinear optical effects. The \textit{Linear-Optic HPS} relies on the interference of two frequency-displaced WCSs in a Hong-Ou-Mandel (HOM) interferometer \cite{HOMPRL1987}. The interference pattern exhibits \textit{anti-bunching peaks} with increased correlation between the interferometer's output spatio-temporal modes. The local detection of a photon in one of these modes enables the output of the source leading to the announcement of a squeezed photon-number pulse. The process is based only on linear optics with no phase-matching restrictions, thus making the proposal freely-tunable over a 100-nm wide spectral range on the telecom bands without performance loss.

The paper is divided as follows. Section \ref{chap2} presents the physical principle driving the proposed source and the model describing the photon-number statistics. Section \ref{chap3} describes the experimental setup used for generating the frequency-displaced WCS and the source implementation. The Hanbury-Brown and Twiss analysis, a simulation of a QKD link, and the results are presented in Section \ref{chap4}. Section \ref{chap5} presents the final remarks and the conclusion.

\section{Two-photon interference with weak coherent states}
\label{chap2}

It has been formerly shown that the incidence of two indistinguishable single photons in a beam splitter (BS) leads to the photon-bunching effect when the temporal modes are matched \cite{HOMPRL1987}. A quantum beat pattern is expected when the photons have different frequencies  \cite{OuPRL1988} and the effect can be observed in temporal modes if their coherence time is high enough \cite{LegeroPRL04,LegeroAPB03}. The optical beat note corresponds to the frequency displacement between the photons even if independent WCSs are used \cite{ThiagoJOSAB15}.

Consider a symmetric BS with input spatial modes $A$ and $B$ and output spatial modes $C$ and $D$. The model considers that the input states are generated by two independent continuous-wave (CW) optical sources emitting spatial-mode matched parallel-polarized photons with identical average number of photons per time interval. We decompose the input frequency-mismatched WCSs into pairs of Fock states with $M$ and $N$ photons, $\vert M,N\rangle_{A,B}$. The error associated to the simplification of bounding $M+N\leq 3$ is estimated from the Poisson cumulative distribution function, $\sum_{M=0}^3\sum_{N=0}^3 \mu^{M+N}e^{-2\mu}/\left(M!N!\right)$ \cite{KnightBOOK}, as being lower than $1\%$ for $\mu<0.67$.

The case in which any number of photons enter the interferometer through one of the input modes, with vacuum at the other, $\vert M,0 \rangle_{A,B}$ or $\vert 0,N\rangle_{A,B}$, has a straightforward solution: a random distribution of photons between the two output spatial modes. We focus, therefore, on the case of two frequency-displaced single photons entering the HOM interferometer, one in each of the input modes. This main result is then used to compose the other non-trivial cases of one- and two-photon states entering each input mode $\vert 1,2\rangle_{A,B}$, $\vert 2,1\rangle_{A,B}$, and $\vert 2,2\rangle_{A,B}$. The possible outcomes of the interferometer are then combined and weighted by the probability of occurrence of the correspondent input states. This information consists on the source's output statistics, once the probability of a heralding event is considered. 

\subsection{Theoretical Model for the Linear-Optic HPS}
The transfer function for the symmetric BS can be described using creation operators \cite{KnightBOOK}:
\begin{align}
\begin{aligned}
a_A^{\dagger} &= \left(ja_C^{\dagger}+a_D^{\dagger}\right)/\sqrt{2} \\
a_B^{\dagger} &= \left(a_C^{\dagger}+ja_D^{\dagger}\right)/\sqrt{2}
\end{aligned}
\end{align}

Combining the description of the wave-packet of each input photon $\xi_{A,B}\left(t\right) = \varepsilon_{A,B}\left(t\right)e^{-j\phi_{A,B}}$ to the creation operators, we write electric field operators that can be associated to the input modes of the BS \cite{LegeroAPB03}:
\begin{align}
\begin{aligned}
 E_{A,B}^+\left(t\right) &= \xi_{A,B}\left(t\right)a_{A,B} \\
 E_{A,B}^-\left(t\right) &= \xi^\ast_{A,B}\left(t\right)a^\dagger_{A,B}
\end{aligned}
\end{align}
 
The BS's output spatio-temporal modes can, therefore, be written as a function of the field operators \cite{ThiagoJOSAB15,LegeroAPB03}
\begin{align}
\begin{aligned}
E_C^{+}\left(t\right) &= \left[-jE_A^{+}\left(t\right)+E_B^{+}\left(t\right)\right]/\sqrt{2} \\
E_D^{+}\left(t\right) &= \left[E_A^{+}\left(t\right)-jE_B^{+}\left(t\right)\right]/\sqrt{2}
\end{aligned}
\end{align}

The case of interest is that of two single photons entering the BS, one at each input mode, which can be written in terms of creation operators applied to vacuum as $\vert \text{in}\rangle_{A,B} = \vert 1,1\rangle_{A,B} = a^\dagger_A a^\dagger_B\vert 0,0\rangle_{A,B}$. We adopt the notation $P^{M,N_{\text{in}}}_{R,S_{\text{out}}}$ for the conditional probability of finding $R$ and $S$ photons at the output modes $C$ and $D$, respectively, given that we had $M$ and $N$ photons at the input modes $A$ and $B$, respectively.

The probability of finding one photon in $C$ at $t = \tau_0$ and the other photon in $D$ at $t = \tau_0+\tau$ can be computed as
\begin{widetext}
\begin{align}
\begin{aligned}
P^{1,1_{\text{in}}}_{1,1_{\text{out}}}\left(\tau_0,\tau\right) &= \langle 0\vert a_A a_B E_C^- \left(\tau_0\right) E_D^- \left(\tau_0+\tau\right)E_C^+ \left(\tau_0\right) E_D^+ \left(\tau_0+\tau\right) a_A^\dagger a_B^\dagger \vert 0\rangle \\
&\text{resulting in}\\
P^{1,1_{\text{in}}}_{1,1_{\text{out}}}\left(\tau_0,\tau\right) &=
\frac{1}{4} \left\vert\xi_A (\tau_0+\tau) \xi_B (\tau_0)-\xi_A (\tau_0) \xi_B (\tau_0+\tau)\right\vert ^2
\end{aligned}
\label{PhInterf}
\end{align}
\end{widetext}

Eq. \ref{PhInterf} can be solved considering gaussian-shaped wave-packets with identical half-width at 1/e, $\sigma$, and well-defined angular frequency modes $\omega_A$ and $\omega_B$ ($\omega=\omega_A/2+\omega_B/2$). The frequency difference $\Delta=\omega_B-\omega_A$ is fixed and the wave-packets can be expressed as 
\begin{align}
\begin{aligned}
\xi_A\left(t\right) &= \frac{1}{\sqrt[4]{\pi\sigma^2}}e^{-\left(t-\delta\tau
/2\right)^2/(2\sigma^2)}e^{-j\left(\omega-\Delta/2\right)t} \\
\xi_B\left(t\right) &= \frac{1}{\sqrt[4]{\pi\sigma^2}}e^{-\left(t+\delta\tau
/2\right)^2/(2\sigma^2)}e^{-j\left(\omega+\Delta/2\right)t}
\end{aligned}
\end{align}
where $\delta\tau$ is the relative delay between the wave-packets.

Integration from $-\infty$ to $\infty$ over $\tau_0$ (as we are only interested in the effective time delay $\tau$ between the wave packets) and over $\delta\tau$ (to take into account the CW nature of the optical sources) yields \cite{ThiagoJOSAB15}
\begin{equation}
P^{1,1_{\text{in}}}_{1,1_{\text{out}}} = \frac{1}{2}\left(1-\beta\right)
\label{1ph11}
\end{equation}
with $\beta = e^{-\frac{\tau^2}{2\sigma^2}}\cos\left(\tau\Delta\right)$. Due to normalization and symmetry arguments, the probability of finding two photons in one output spatial mode and vacuum in the other is found to be
\begin{equation}
P^{1,1_{\text{in}}}_{2,0_{\text{out}}} = P^{1,1_{\text{in}}}_{0,2_{\text{out}}} = \frac{1}{4}\left(1+\beta\right)
\label{1ph20}
\end{equation}

The solution considering a single photon entering one of the input ports of the beam splitter with vacuum entering the other, $\vert 1,0\rangle_{in}$ and $\vert 0,1\rangle_{in}$, is straightforward: 
\begin{equation}
P^{1,0_{\text{in}}}_{1,0_{\text{out}}} = P^{1,0_{\text{in}}}_{0,1_{\text{out}}} = P^{0,1_{\text{in}}}_{1,0_{\text{out}}} = P^{0,1_{\text{in}}}_{0,1_{\text{out}}} = 1/2
\label{1ph10}
\end{equation}

The same analysis applies to the case in which either two (Eq.\ref{2ph}) or three photons (Eq.\ref{3ph}) enter the same input port and vacuum enters the other
\begin{align}
\begin{aligned}
P^{2,0_{\text{in}}}_{2,0_{\text{out}}} = P^{2,0_{\text{in}}}_{0,2_{\text{out}}} = P^{0,2_{\text{in}}}_{2,0_{\text{out}}} = P^{0,2_{\text{in}}}_{0,2_{\text{out}}} &= 1/4 \\
P^{2,0_{\text{in}}}_{1,1_{\text{out}}} = P^{0,2}_{1,1_{\text{out}}} &= 1/2
\end{aligned}
\label{2ph}
\end{align}
\begin{align}
\begin{aligned}
P^{3,0_{\text{in}}}_{3,0_{\text{out}}} &= P^{3,0_{\text{in}}}_{0,3_{\text{out}}} = P^{0,3_{\text{in}}}_{3,0_{\text{out}}} = P^{0,3_{\text{in}}}_{0,3_{\text{out}}} = 1/8 \\
P^{3,0_{\text{in}}}_{2,1_{\text{out}}} &= P^{3,0_{\text{in}}}_{1,2_{\text{out}}} = P^{0,3_{\text{in}}}_{2,1_{\text{out}}} = P^{0,3_{\text{in}}}_{1,2_{\text{out}}} = 3/8
\end{aligned}
\label{3ph}
\end{align}

The cases $\vert 2,1\rangle_{\text{in}}$ and $\vert 1,2\rangle_{\text{in}}$ are modeled as a superposition of a pair of single photons, one in each input port, and an independent photon in either one of the input ports. 
\begin{align}
\begin{aligned}
P^{2,1_{\text{in}}}_{3,0_{\text{out}}} &= P^{2,1_{\text{in}}}_{0,3_{\text{out}}} = P^{1,2_{\text{in}}}_{3,0_{\text{out}}} = P^{1,2_{\text{in}}}_{0,3_{\text{out}}} = \frac{1}{8}\left(1+\beta\right) \\
P^{2,1_{\text{in}}}_{2,1_{\text{out}}} &= P^{2,1_{\text{in}}}_{1,2_{\text{out}}} = P^{1,2_{\text{in}}}_{2,1_{\text{out}}} = P^{1,2_{\text{in}}}_{1,2_{\text{out}}} = \frac{1}{8}\left(3-\beta\right)
\end{aligned}
\label{3ph21}
\end{align}

The final probabilities for each output state in modes $C$ and $D$ are obtained by grouping the conditional probabilities of each output pair, Eqs. (\ref{1ph11}-\ref{3ph21}), weighted by the probabilities of occurrence of each pair of input Fock states. The latter are given, for WCSs, by the Poisson distribution $P\left(M,N|\mu\right)_{A,B}=\frac{\mu^{M+N}}{M!N!}e^{-2\mu}$ \cite{KnightBOOK}. The general form of the probability of occurrence of the output state $\vert R,S\rangle_{C,D}$ when the BS is fed by frequency-displaced WCSs is, thus, given by
\begin{equation}
P_{R,S} = \sum_{M=0}^3\sum_{N=0}^3 P^{M,N}_{R,S}P_{M,N},
\left\{ \begin{array}{l}
R,S,M,N \in \mathbb{N}\\
R+S\leq 3\\
M+N\leq 3
\end{array}\right.
\label{GeneralOutputProb}
\end{equation}

\subsection{Statistics of the Linear-Optic HPS}

The output states of the source are conditioned to the detection of one -- or more than one -- photon at the spatial mode $C$, the heralder arm. We thus normalize the emission statistics of the source using those cases when at least one photon heralds the output mode, which is given by
\begin{equation}
P_{T} = \sum_{R=1}^3\sum_{S=0}^2 \eta_{R} P_{R,S}, 
\left\{ \begin{array}{l}
R,S \in \mathbb{N}\\
R+S\leq 3
\end{array}\right.
\end{equation}

Therefore, the probabilities of the three distinct events of heralding vacuum $P_v$, single- $P_s$ or multi-photon $P_m$ pulses at the spatial mode $D$ (disregarding dark counts) are given by
\begin{equation}
P_v = \left(\eta_1 P_{1,0}+\eta_2 P_{2,0}+\eta_3 P_{3,0}\right)/P_{T}
\label{Pvacuum}
\end{equation}
\begin{equation}
P_m = \eta_1 P_{1,2}/P_{T}
\label{Pmulti}
\end{equation}
\begin{equation}
P_s = \left(\eta_1 P_{1,1}+\eta_2 P_{2,1}\right)/P_{T}
\label{Psingle}
\end{equation}
where $\eta_i=1-\left(1-\eta_C\right)^i$ is the $i$-photon detection efficiency of the heralder detector of overall detection efficiency $\eta_C$. With $\alpha = \exp\left(-2\mu\right)$, Eqs. (\ref{Pvacuum}-\ref{Psingle}) can be expressed as 

\begin{widetext}
\begin{align}
\begin{aligned}
P_v &= \frac{
24+
\mu  \left[ 24-12\eta_C+\left(12-6\eta_C\right)\beta\right]+
\mu^2\left[ 12-12\eta_C+4\eta_C^2+\left(9-9\eta_C+3\eta_C^2\right)\beta\right]
}{
24+
\mu  \left( 48-12\eta_C-6\eta_C\beta\right)+
\mu^2\left[48-24\eta_C+4\eta_C^2+\left(3\eta_C^2-6\eta_C\right)\beta\right]
} \\
P_m &= \frac{
\mu^2\left(12-3\beta\right)
}{
24+
\mu  \left( 48-12\eta_C-6\eta_C\beta\right)+
\mu^2\left[48-24\eta_C+4\eta_C^2+\left(3\eta_C^2-6\eta_C\right)\beta\right]
} \\
P_s &= \frac{
\mu\left(24-12\beta\right)+
\mu^2\left[24-12\eta_C+\left(3\eta_C-6\right)\beta\right]
}{
24+
\mu  \left( 48-12\eta_C-6\eta_C\beta\right)+
\mu^2\left[48-24\eta_C+4\eta_C^2+\left(3\eta_C^2-6\eta_C\right)\beta\right]
}
\label{EqPs}
\end{aligned}
\end{align}
\end{widetext}

\section{Experimental setup}
\label{chap3}

\begin{figure}[ht]
\center
\includegraphics[trim=0cm 5cm 1.5cm 0cm, clip=true, width=0.45\textwidth]{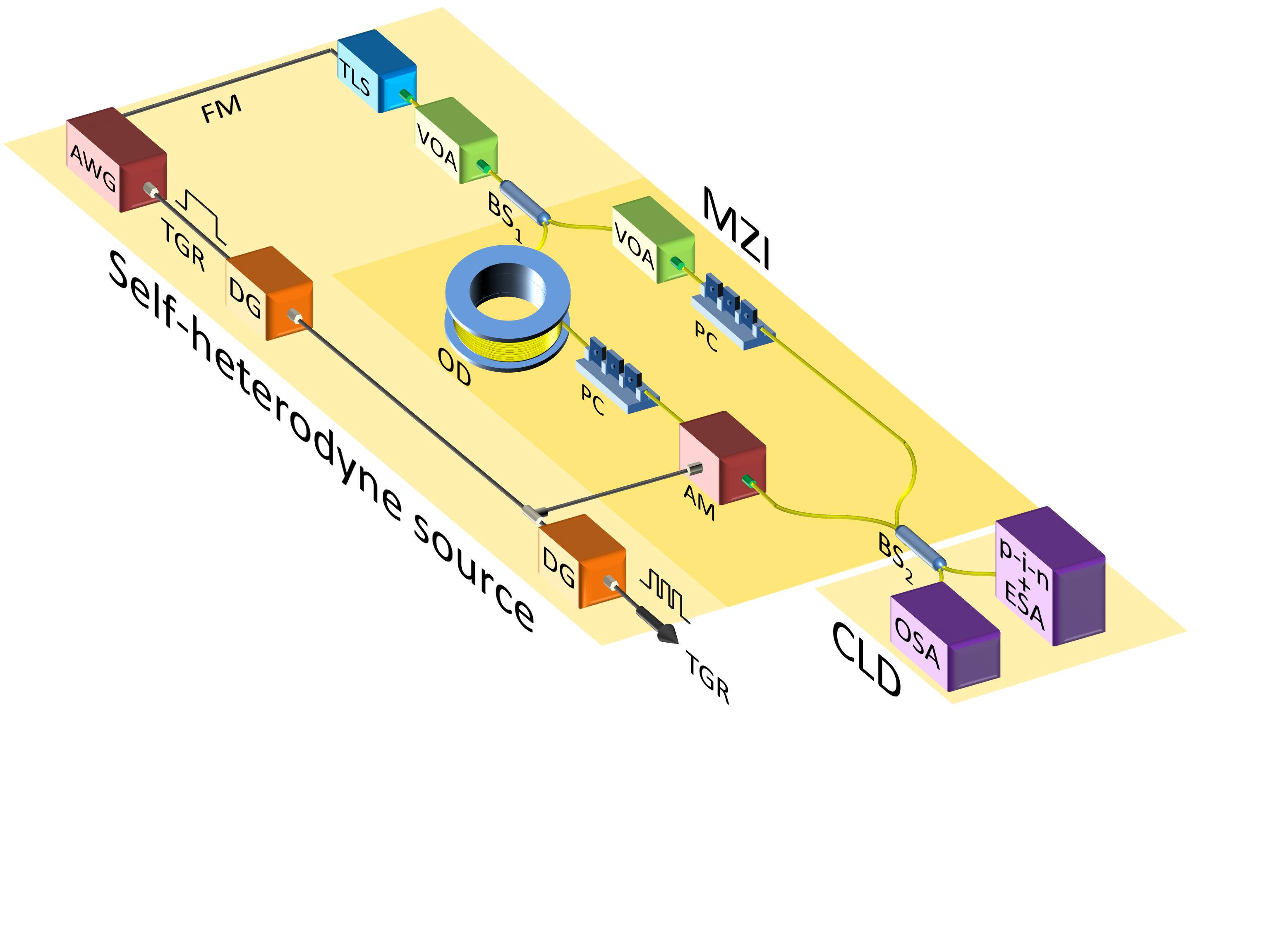}
\caption{Preparation of frequency-displaced WCSs with a self-heterodyne setup. 
AWG: arbitrary waveform generator;
FM: frequency modulation;
TGR: trigger signal;
TLS: tunable laser source;
VOA: variable optical attenuator;
BS: beam splitter;
OD: optical delay;
DG: delay generator;
PC: polarization controller;
AM: amplitude modulator;
CLD: classical-level detection unit.
}
\label{Fig2}
\end{figure}

We report an \textit{all-fiber} implementation of the Linear-Optic HPS source. Two frequency-displaced WCSs create an optical beat note whose anti-bunching peaks are explored for heralding Sub-Poisson photon number pulses. The central wavelength of the source is chosen to be 1545.90 nm, with spectral tunability over 100 nm around the telecom C band. The self-heterodyne approach, \cite{ThiagoJOSAB15}, depicted in Fig. \ref{Fig2}, generates the two frequency-displaced WCSs up from a tunable laser source. The technique avoids stability issues regarding both optical power and relative frequency drift. 

\begin{figure}[ht]
\center
\includegraphics[trim=0cm 0cm 0cm 0cm, clip=true, width=0.45\textwidth]{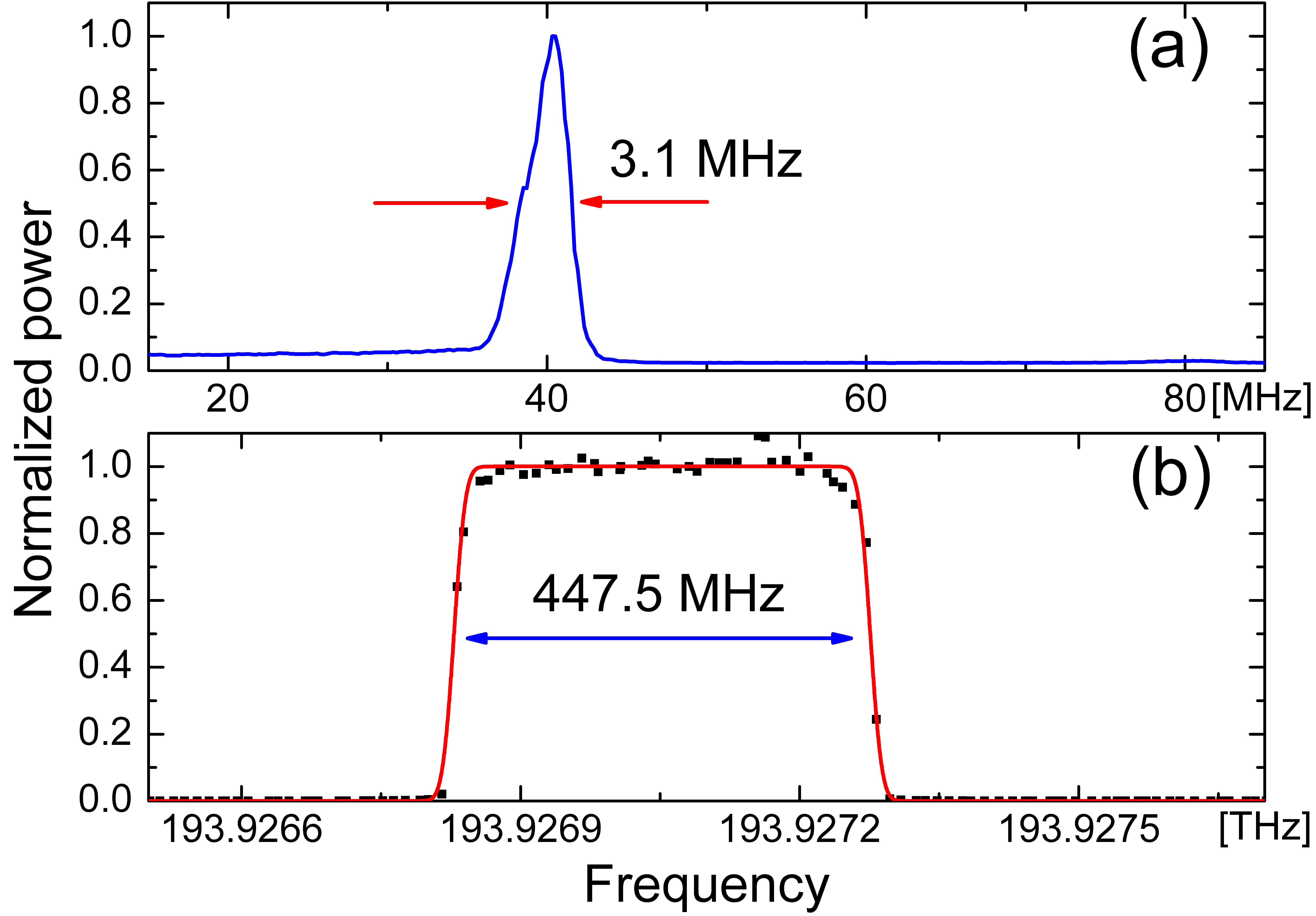}
\caption{Measurement of bright versions of the WCSs at the classical-level detection unit: (a) beat spectrum and (b) optical spectrum.
}
\label{Fig3}
\end{figure}

The attenuated CW signal of an external cavity laser diode is frequency modulated with a triangular wave of period $T = 322.6\mu$s and sent into a balanced Mach-Zehnder Interferometer-like setup (MZI). An optical switch at one of the MZI's arms selects the photons with constant frequency offset relative to the other arm. The optimized duration of the optical switch pulse was experimentally determined to be of $30\mu$s -- 9.3 $\%$ of the triangular wave period. This value yields a classical optical beat note of 40 MHz with a 3.1 MHz linewidth verified by measuring the bright-light version of the optical beat note with a p-i-n photodiode and an electrical spectrum analyzer at the output of the MZI (Fig. \ref{Fig3}-a).

The total frequency sweep that characterizes the Linear-Optic HPS output linewidth is experimentally determined with a high resolution Optical Spectrum Analyzer connected to the source's output (Fig. \ref{Fig3}-b) as 447.5 MHz. The measured linewidth is square-shaped due to the linear sweep of the frequency modulation. The correspondent coherence time of 2.2 ns is comparable to ultra-narrow filtered SPDC-based HPSs \cite{ClausenNJP14}.

\begin{figure}[ht]
\center
\includegraphics[trim=0cm 2.8cm 2.7cm 0cm, clip=true, width=0.45\textwidth]{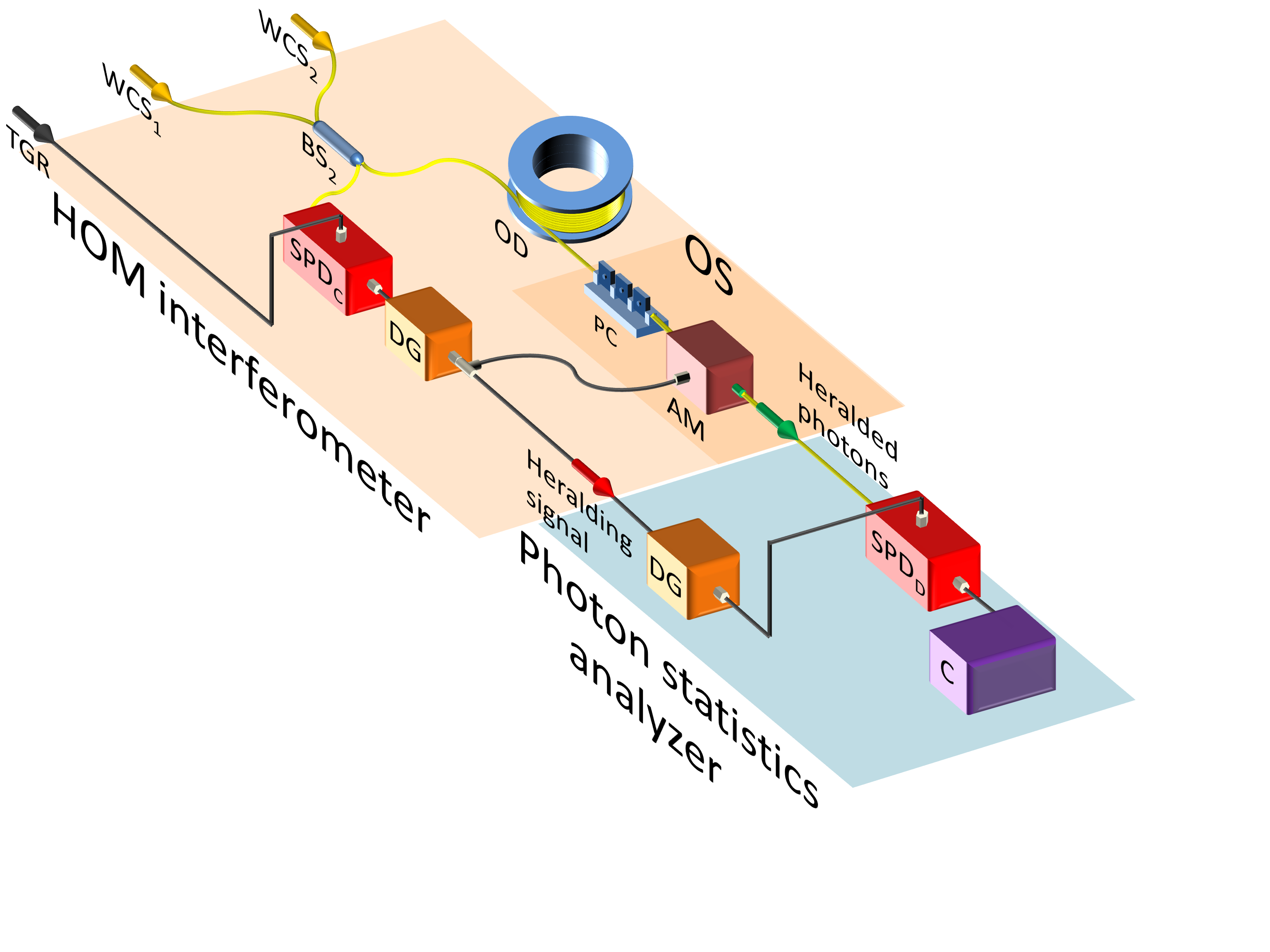}
\caption{HOM interferometer with overlapping WCSs and photon statistics analyzer.
OS: optical switch;
C: pulse counter.
}
\label{Fig4}
\end{figure}

The output of the MZI is converted into an HOM interferometer to build up the Linear-Optic HPS, which can be time-tuned so that the local detection at the heralder mode enables the source's optical output. In order to determine the operational point of the source, the interference pattern is measured with a photon statistics analyzer -- composed of a remote detector triggered by a delayed version of the heralding signal -- connected to its output, as depicted in Fig. \ref{Fig4}. All detectors are InGaAs avalanche photodiode single-photon detectors (SPD) operating in gated-Geiger mode with 2.5 ns temporal gate width, 15$\%$ detection efficiency and 10 $\mu$s deadtime.

\begin{figure}[ht]
\center
\includegraphics[trim=0cm 0cm 0cm 0cm, clip=true, width=0.45\textwidth]{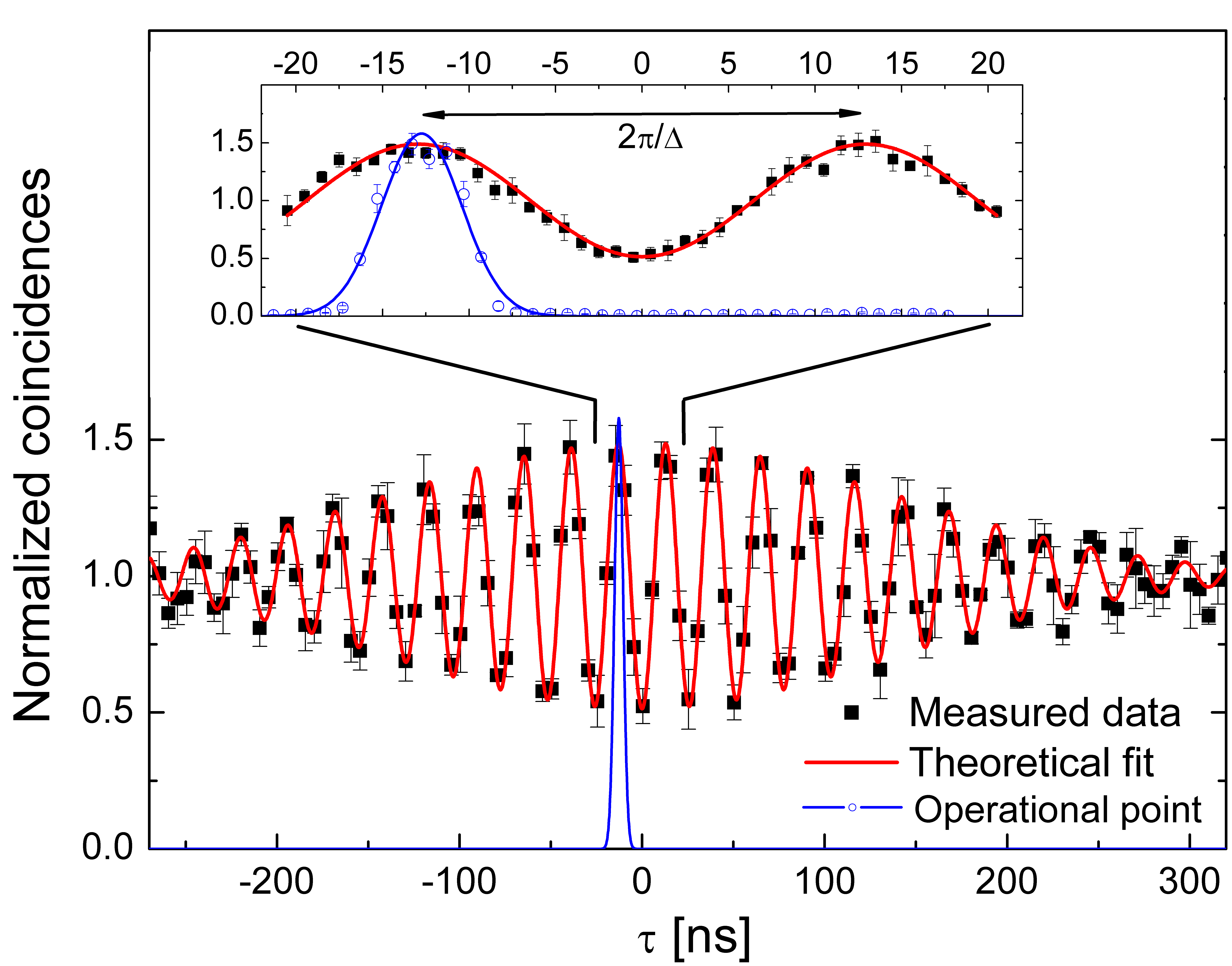}
\caption{Interference pattern measured with the photon statistics analyzer for the 40MHz-displaced WCSs with the optical switch matched to the operational point (blue) and synchronously swept with the detector (black squares). The theoretical model is fit to data (red line).
}
\label{Fig5}
\end{figure}

Figure \ref{Fig5} depicts the beat pattern of the HOM interferometer when both the optical switch and the heralded pulse's detection are synchronously scanned with respect to the heralding signal. The pattern is characterized by the mutual coherence time of the sources and by their relative frequency mismatch \cite{LegeroAPB03,ThiagoJOSAB15}. The theoretical prediction for the normalized coincidence counts \cite{ThiagoJOSAB15}
\begin{equation}
C_{coinc} = \frac{1}{2}\left(2-e^{-\frac{\tau^2}{2\sigma^2}}\cos\left(\tau\Delta\right)\right)
\end{equation}
is fit to data.

The operational point for the Linear-Optic HPS is also shown in Fig. \ref{Fig5}, which is obtained when the optical switch is matched and fixed to $\tau = \pm\pi/\Delta$ -- the anti-bunching peak. Under this configuration, only the photons enabled by the gate opened at the output optical switch -- and thus conditioned to the local detection at the heralder detector -- are emitted. Even though the overall output of our source exhibits a coherence time of 2.2 ns, the optical switch's gate width may be a limiting factor for the coherence time. Given the $\sim5$ ns gate width shown in Fig. \ref{Fig5}, however, this does not impact on the photon's coherence. It is worthy noting that, in a single-photon interference context, the linewidth is limited to around $200$ MHz according to the Fourier Transform of the output pulse width.

\section{Results}
\label{chap4}

\subsection{The Second-order correlation function at zero time}

Two distinct values of the relative temporal mode delay, $\tau$, are of importance in analyzing the source's statistics: $\tau$ greater than the mutual coherence time of the WCSs ($\tau_{coh}$); and $\tau=\pm\pi/\Delta$, the operational point of the source which corresponds to the highest anti-bunching peak of the interference pattern. In the first case, the interference vanishes since the states are \textit{fully distinguishable}. Any attempt to herald such states will fall into the Poisson-like photon-number statistics of a faint laser source.

In the anti-bunching peak case, however, coincident counts can reach up to $150\%$ of the distinguishable case -- see Fig. \ref{Fig5}. The normalized heralding probabilities of vacuum, multi-photon, and single-photon pulses are presented below as a function of $\mu$ and of the heralder SPD's efficiency ($\eta_C$) where the value of $\beta$ was approximated to $-1$ (anti-bunching condition) in Eq. \ref{EqPs}:
\begin{equation}
P_v = \frac{8+\mu  \left( 4-2\eta_C\right)+\mu^2\left( 1-\eta_C+\eta_C^2/3\right)}
{8+\mu  \left( 16-2\eta_C\right)+\mu^2\left(16-6\eta_C+\eta_C^2/3\right)}
\end{equation}
\begin{equation}
P_m = \frac{5\mu^2}
{8+\mu  \left( 16-2\eta_C\right)+\mu^2\left(16-6\eta_C+\eta_C^2/3\right)}
\end{equation}
\begin{equation}
P_s = \frac{12\mu+\mu^2\left( 10-5\eta_C\right)}
{8+\mu  \left( 16-2\eta_C\right)+\mu^2\left(16-6\eta_C+\eta_C^2/3\right)}
\end{equation}

In Table \ref{table:probs}, we present these normalized heralding probabilities considering $\mu << 1$ and $\eta_C = 1$ compared to the probability of finding similar pulses in a faint laser source. Even though the multi-photon emission probability is slightly higher, the emission of vacuum is relatively suppressed, and the single-photon pulses occur with higher probability. This characterizes the narrowing of the photon-number distribution of the squeezed WCS at the output of the source \cite{DavidovichRMP1996}.

\begin{table}[ht]
\centering
\caption[caption]{Normalized Probabilities ($\mu<<1$,$\eta_C=1$)}
\renewcommand{\arraystretch}{1.4}
\begin{tabular}{c | c c c}
\hline\hline
Pulse 					& Linear-Optic HPS							& Faint laser source					& Ratio    \\
\hline
Vacuum         	& $1-\frac{3}{2}\mu+\frac{15}{8}\mu^2$	& $1-\mu+\frac{1}{2}\mu^2$ 	&	$1-\frac{1}{2}\mu$			\\				
Multi-Photon    	& $\frac{5}{8}\mu^2$					& $\frac{1}{2}\mu^2$ 				&	$\frac{5}{4}-\frac{4}{3}\mu$	\\				
Single-Photon   	& $\frac{3}{2}\mu-2\mu^2$				& $\mu-\mu^2$ 			&	$\frac{3}{2}-\frac{1}{2}\mu$		\\[1ex]		
\hline
\end{tabular}
\label{table:probs}
\end{table}

\begin{figure}[ht]
\center
\includegraphics[trim = 0cm 9.3cm 10.5cm 0cm, clip=true, width=0.45\textwidth]{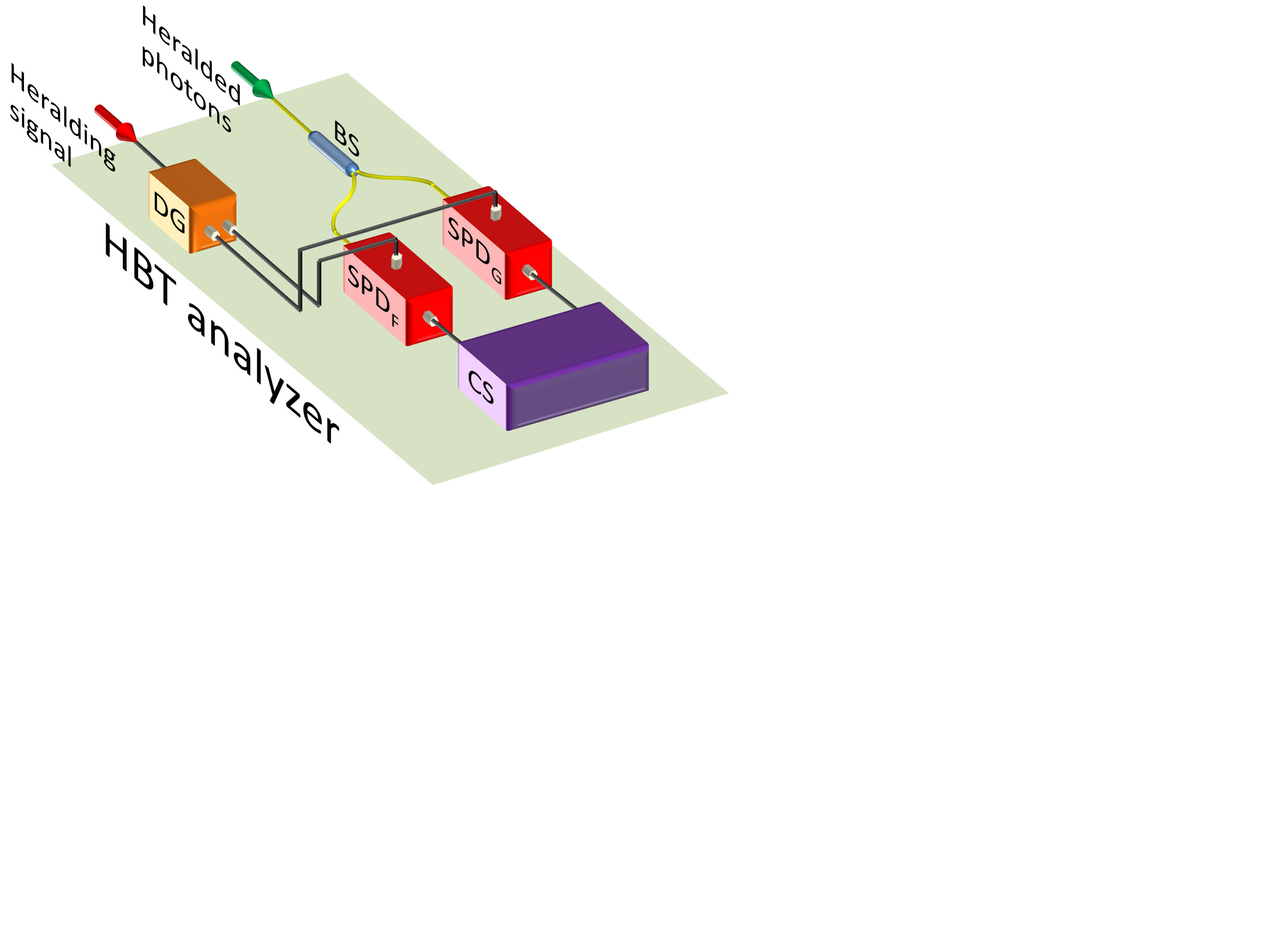}
\caption{Experimental setup for the HBT Analyzer.}
\label{Fig6}
\end{figure}

The second-order correlation function parameter $g^2\left(\tau\right)$ at zero time determines whether the source produces photons following sub- $\left(< 1\right)$, super- $\left( > 1\right)$, or Poisson-like $\left(= 1\right)$ statistics \cite{FaselNJP04,WangPRL08}. The experimental values are obtained with a Hanbury-Brown and Twiss (HBT) Analyzer set at the output of the source. The HBT setup consists of a symmetrical BS connected to the heralded mode, an SPD connected to each of its output ports triggered by the heralding signal, and a coincidence station (CS) (Fig. \ref{Fig6}).

The parameter $g^2(0)$ is experimentally defined as \cite{FaselNJP04}
\begin{equation}
g^2(0) = Q_{FG}/\left(Q_F Q_G\right)
\end{equation}
where, $Q_F$, $Q_G$, and $Q_{FG}$ are, respectively, the probabilities of an event being registered at SPD F, SPD G, and in coincidence between them. These values can be computed from the probability of an $i$-photon state being heralded (the source statistics) \textit{and} detected in the desired detector (the single-photon detection probability given by $\eta_F$ and $\eta_G$). For negligible dark count probability, a detection event can be triggered by a single- or multi-photon pulse, so we can write
\begin{equation}
Q_{F,G} = \frac{P_s\eta_{F,G}}{2}+\left(\eta_{F,G}-\frac{\eta_{F,G}^2}{4}\right)P_m
\end{equation}

Given our suppositions, coincident detections are triggered by half the multi-photon probability events, resulting in
\begin{equation}
Q_{FG} = \frac{\eta_F\eta_G}{P_m/2}
\end{equation}
The $g^2\left(0\right)$ parameter is then re-written as a function of the photon statistics of the Linear-Optic HPS and the detection efficiency values as

\begin{widetext}
\begin{equation}
\begin{split}
g^2\left(0\right) = \frac{P_m}{P_s^2/2+\left(2-\eta_F/4-\eta_G/4+\eta_F\eta_G/8\right)P_m^2+\left(2-\eta_F/4-\eta_G/4\right)P_sP_m}
\end{split}
\end{equation}
\end{widetext}

\begin{figure}[ht]
\center
\includegraphics[trim = 0cm 0cm 0cm 0cm, clip=true, width=0.45\textwidth]{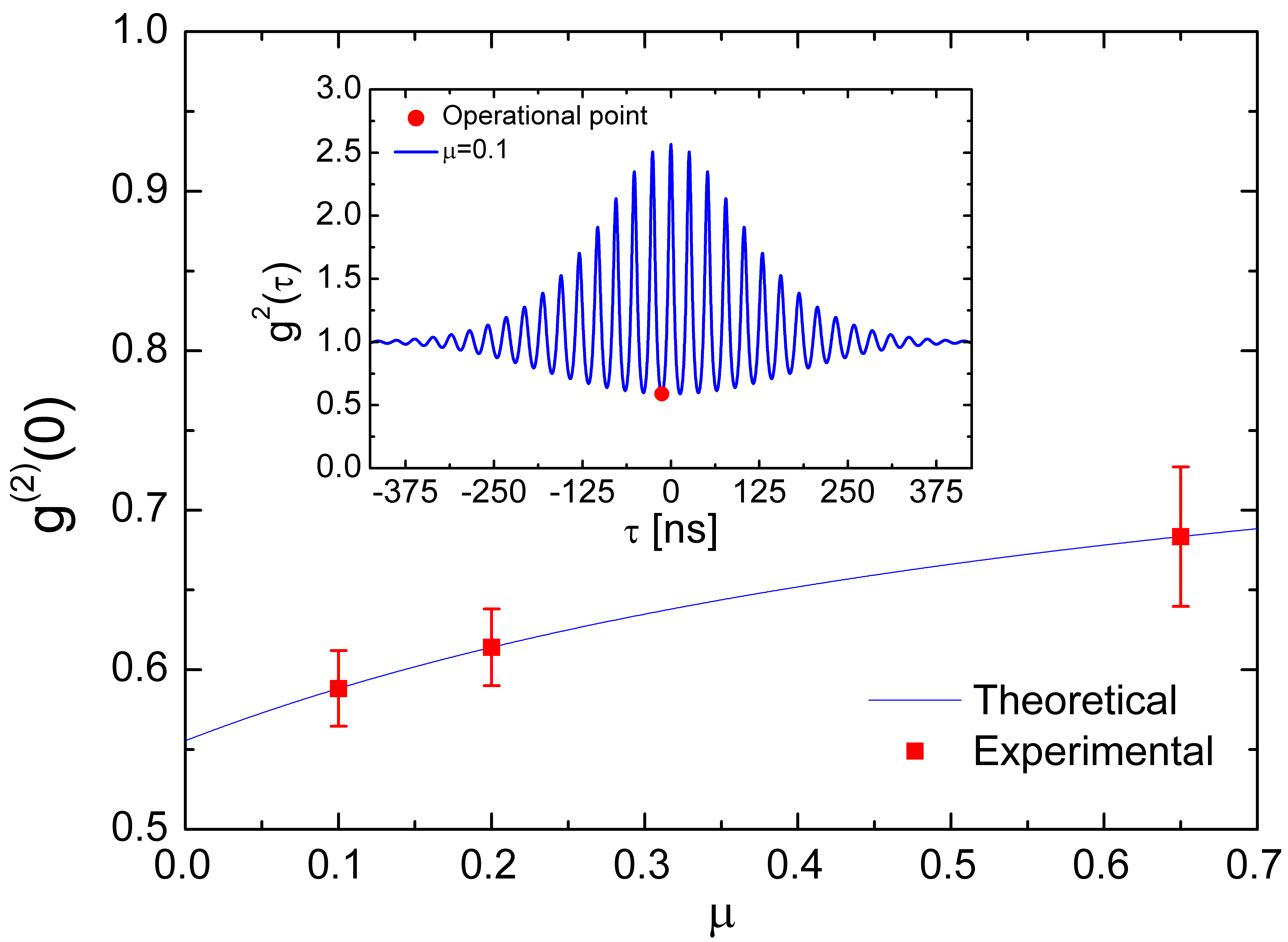}
\caption{Theoretical values and experimental results for $g^2\left(0\right)$ at the operational point. Inset shows the $g^2\left(\tau\right)$ function.}
\label{Fig7} 
\end{figure}

From the modeled single and multi-photon probabilities, we find that $g^2(0)$ is 0.56 for the \textit{anti-bunching} and 1 for the distinguishable cases. Fig. \ref{Fig7} presents the predicted values for $g^2\left(0\right)$ and the experimental values. The uncertainty in the experimental value found for the $g^{(2)}(0)$ parameter comes from the fit of the experimental parameters (detection efficiency and average number of photons per pulse) to the measured gain (obtained with the setup shown in Fig. \ref{Fig6}) and from the truncation error on the model due to the three-photons approach.

The inset exhibits the predicted values for $g^2\left(\tau\right)$ considering $\mu = 0.1$ and the detectors efficiency $\eta_{H,F,G} = 0.15$. The agreement between experimental and theoretical results, shown in the main figure, validates the model and indicates the sub-poissonian character of the source due to $g^2\left(0\right) < 1$.

\subsection{Simulation of a quantum key distribution link}

The performance of a QKD system is directly dependent on the photon statistics of the optical source. The secret key generation probability and the maximum achievable link distance are analyzed below for our Linear-Optic HPS, for a faint laser source, and for an SPDC-based HPS, following both the GLLP security analysis \cite{GottesmanQIC04} and the decoy states approach \cite{HwangPRL03,MaPRL05,WangPRL05}. The parameters $P_0$, $P_1$ and $P_2$ are the probability of Alice sending vacuum, single- or multi-photon pulses for each kind of source considered. 

The yield is the conditional probability of detection at Bob, given that Alice has sent an $i$-photon pulse. This is given by $Y_i\approx P_d + \eta_i$, where $P_d$ is the detector dark count probability and $\eta_i=1-\left(1-\eta\right)^i$. The overall transmittance $\eta$ is composed by the product of the efficiency  of Bob's detection devices ($\eta_{Bob}$) and the link loss ($10^{-\alpha L/10}$) -- where $\alpha$ is the attenuation coefficient (dB/km) of the fiber with length $L$ (km) at the operation wavelength. The gain of the $i$-photon state is given by $Q_i=P_i Y_i$, and the overall gain is obtained by summing over the contribution of all states $Q_{\mu} = \sum_{i=0}^\infty Q_i$.

The error probability of the \textit{i}-photon state is given by $e_i=\left(e_0 Y_0+e_{opt}\eta_i\right)/Y_i$, where $e_0$=1/2 is the probability of a dark count to occur at the wrong detector and $e_{opt}$ corresponds to the optical misalignment of Bob's apparatus. The overall error is given by $E_\mu=\frac{1}{Q_{\mu}}\sum_{i=0}^\infty e_i Q_i$.

The secret key generation probability is computed as

\begin{equation}
R\geq q\left\{Q_1 \left[1-H_2\left(e_1\right)\right]-Q_\mu H_2\left(E_\mu\right)f\left(E_\mu\right)\right\}
\end{equation}
where $H_2\left(x\right)$ is the Shannon binary entropy, $f(x)$ is the inefficiency of the error correction (1.16 here), and q=1/2 appears due to the bases matching probability.

Considering the GLLP security analysis \cite{GottesmanQIC04}, the gain of single-photon pulses is lower bounded by

\begin{equation}
Q_1=Q_\mu-\sum_{i=2}^\infty P_i\approx Q_\mu-P_{m}\left(\mu\right)
\end{equation}
This assumption is the pessimistic one where all multi-photon pulses sent by Alice are eavesdropped. In this analysis, the single-photon error is given by
\begin{equation}
e_1=E_\mu Q_\mu/Q_1
\end{equation}

When using the decoy states method, the values of $Q_1$ and $e_1$ can be estimated without the pessimistic assumption. This is possible due to the random choice of intensities sent by Alice. The $i$-photon yield does not depends on Alice's choice, so Eve cannot predict nor fake the output of Alice's source. Furthermore, due to these tighter bounds, Alice can usually use higher intensity values without jeopardizing the system security.

The simulation parameters were extracted from \cite{GobbyAPL04}: $\alpha$=0.21 dB/km; $\eta_{Bob}$=0.045; $P_{d}$=0.85$\times$10$^{-6}$; $e_{opt}$=0.033. The value of $\mu$ was optimized at each link distance for the faint laser source and Linear-Optic HPS. The efficiency of the heralder detector of the Linear-Optic HPS is 15\%. The probabilities for the SPDC-based HPS were obtained from \cite{PomaricoOPEX12}: $P_1$=0.42 and $g^2\left(0\right)$=0.018, with $P_2=g^2\left(0\right)P_1^2/2$ \cite{FaselNJP04}. 

\begin{figure}[ht]
\includegraphics[trim=0cm 0cm 0cm 0cm, clip=true, width=0.45\textwidth]{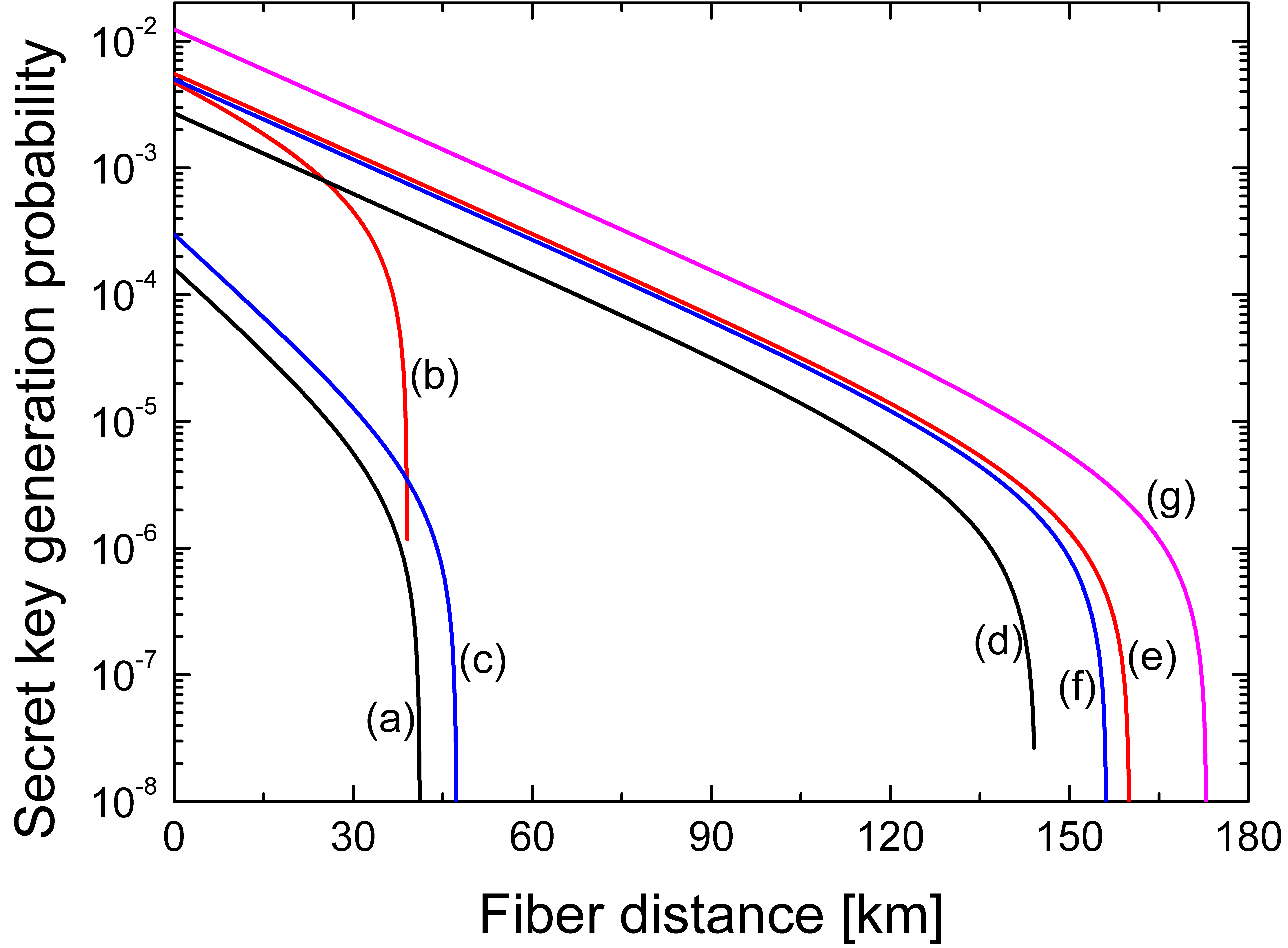}
\caption{Simulation of the secret key generation probability for a BB84-based QKD system considering: GLLP analysis with (a) faint laser source, (b) SPDC-based HPS, and (c) Linear-Optic HPS; decoy states method with (d) faint laser source, (e) SPDC-based HPS, and (f) Linear-Optic HPS; and (g) a true single-photon source.}
\label{Fig8}
\end{figure}

The results for the secret key generation probability obtained as a function of the link length for different sources are shown in Fig. \ref{Fig8}. 

\subsection{Discussion}

The reported Linear-Optic HPS, when conditioned to the temporal selection of pulses at the operational point, exhibits improved statistics when compared to that of coherent states. The output states may be regarded as squeezed photon-number distribution weak coherent states since the second-order correlation function at zero time is less than one \cite{KnightBOOK, DavidovichRMP1996}.

The present QKD link simulation reveals that, under GLLP analysis, the proposed source increases the maximum achievable distance of a QKD link with a factor of $1.15$ when compared to both the faint laser source and SPDC-based HPS. Considering short links, the secret key generation probability of the source is (at least 1.86 times) greater than for the faint laser source, while largely overcome by the SPDC-based HPS (with a factor 16).

Considering the decoy states approach, we found that the proposed source outperforms the faint laser source in secret key generation probability and in distance. Its performance is also highly competitive with state-of-the-art SPDC-based HPS concerning both parameters. The maximum achievable distance is slightly shorter (4 km, corresponding to a length reduction of 2.5\%) and the secret key generation probability presents a small penalty on the simulation. Parameters taken from \cite{NgahLPR15} were also tested with qualitatively similar results not shown in the figure.

\section{Conclusion}
\label{chap5}

We have presented for the first time, to our knowledge, a heralded photon source relying on linear optics and weak coherent states. The interference between the frequency-displaced weak coherent states in an Hong-Ou-Mandel interferometer results in a structured interference pattern that exhibits anti-bunching peaks. The system is time-tuned so that only photon pulses that match the highest of those peaks are heralded. This enhances the correlation between the locally-detected spatio-temporal mode and the heralded mode. Sub-Poissonian photon statistics can be synchronously achieved. The photon statistics of our linear-optic heralded photon source was modeled and an Hanbury-Brown and Twiss analysis assessed its sub-Poisson character, with a second-order correlation parameter down to 0.556.

The performance of the Linear-Optic HPS was evaluated in a simulation of a BB84-like QKD section. The source was compared to a faint laser source and to an SPDC-based heralded photon-source under both the GLLP security analysis and the decoy states approach. The simulation results reveal that the proposed source outperforms the faint laser source and is highly competitive with the SPDC-based HPS.

The present 2 kHz heralding rate of the self-heterodyne scheme can be improved by employing two continuous wave lasers with fixed displaced frequencies. We estimate that the heralding rate can be as high as $1.5$ MHz, a rate comparable to state-of-the-art SPDC HPSs \cite{NgahLPR15}. This value is achieved when we assume a 100 MHz triggering rate for the local detector, with a detection efficiency of 15\%, and CW independent laser sources emitting 0.1 photons per time interval.

The absence of phase-matching restrictions, as linear optics is used, renders the proposed source freely-tunable over a 100-nm wide spectral range in the telecom bands without performance lost. The physical principle could also be explored to assemble a heralded photon source in the visible spectral range, matching free-space quantum communications.

In conclusion, the all-fiber telecom-compatible design, its sub-poissonian character, and the rather simple setup make the Linear-Optic Heralded Photon Source a proper choice for long-distance quantum communication, broadening the plethora of technological resources in the field.

\section{Acknowledgements}
The authors thank G. B. Xavier, G. Svetlichny, L. Davidovich and F. Brussi\`{e}res for enlightening discussions.


\end{document}